\documentclass[fleqn,10pt]{wlscirep}

\usepackage{amssymb}

\title{Particle trapping and conveying using an optical Archimedes' screw}

\author[1]{Barak Hadad $\ddagger$}

\author[1]{Sahar Froim $\ddagger$}

\author[2]{Harel Nagar}

\author[2]{Tamir Admon}

\author[1]{Yaniv Eliezer}

\author[2]{Yael Roichman}

\author[1]{Alon Bahabad*}

\affil[1]{Department of Physical Electronics, School of Electrical Engineering, Fleischman Faculty of Engineering, Tel-Aviv University, Tel-Aviv 69978, Israel}

\affil[2]{School of Chemistry, Raymond and Beverly Sackler Faculty of Exact Sciences, Tel Aviv University, Tel Aviv 69978, Israel}

\affil[$\ddagger$]{Equal contribution}

\affil[*]{alonb@eng.tau.ac.il}

\begin{abstract}
	
Trapping and manipulation of particles using laser beams has become an important tool in diverse fields of research. In recent years, particular interest is given to the problem of conveying optically trapped particles over extended distances either down or upstream the direction of the photons momentum flow. 
Here, we propose and demonstrate experimentally an optical analogue of the famous Archimedes' screw where the rotation of a helical-intensity beam is transferred to the axial motion of optically-trapped micro-meter scale airborne carbon based particles. With this optical screw, particles were easily conveyed with controlled velocity and direction, upstream or downstream the optical flow, over a distance of half a centimeter. Our results offer a very simple optical conveyor that could be adapted to a wide range of optical trapping scenarios.  

\end{abstract}

\begin{document}

\flushbottom

\maketitle

\thispagestyle{empty}

\section*{Introduction}


Optical trapping and translation of particles using focused laser beams has become an important and useful tool in various disciplines such as biology, atomic physics, optics, thermodynamics, and atmospheric sciences \cite{bowman2013optical,daly2015optical,dienerowitz2008optical,grier2003revolution,moffitt2008recent,woerdemann2013advanced}.
For weakly absorbing, sub-wavelength particles, optical trapping and manipulation is based on the interplay of two forces that act on the trapped particles, both of which are caused by radiation pressure \cite{bowman2013optical}. The first force depends on the gradient of the light’s intensity, which attracts the particle towards the focal point of the beam. The second force is the scattering force, which usually pushes the particle downstream with the beam’s direction of propagation (in special arrangements the scattering force can be set to pull particles against the momentum flow \cite{brzobohaty2013experimental}) . These two forces need to counteract each other, in order to obtain a stable optical trap. 
With absorbing particles, the photophoretic force becomes dominant. The photophoretic force is proportional to the particles size and to the light’s intensity. This force is thermal in nature, namely, when light hits one side of the particle that side becomes warmer increasing the pressure of the nearby air molecules. The resulting pressure drop between the warm and cold sides of the particle pushes them downstream, and away from the light. This force tends to repel the absorbing particle from high intensity regions of the optical beam. It has been shown,  that the ratio between the photophoretic forces and the radiation pressure forces, for absorbing sub-wavelength particles, is close to $ 10^4 $ \cite{desyatnikov2009photophoretic}. Therefore, for such particles, radiation pressure forces can be neglected, while the photophretic force governs particle transportation up to a meter length scale \cite{shvedov2010giant}. 
Special interest is devoted to developing optical tractor beams, which allow the transfer of particles against the direction of the beam's propagation, i.e. the direction of the beam's momentum. This poses an interesting challenge since it requires moving a particle against the direction of both photophoretic forces and radiation pressure when a simple Gaussian beam is used. In addition to moving particles upstream, such tractor beams, also known as optical conveyors, are generally required to transport particles in both directions, upstream and downstream, in a controlled manner. 
For weakly absorbing particles, in a  solution, optical conveyors were realized using a superposition of two Bessel beams, while changing the relative phase between them \cite{ruffner2012optical}. This change in phase, shifts the standing-wave pattern of the structured beam, carrying along particles trapped at the standing-wave intensity crests. 
For airborne absorbing particles which are expelled from high intensity regions, the common practice is to use hollowed beams having dark volumes to trap the particle and then move the dark volume \cite{zhang2011trapping,shvedov2011robust}. Alternatively, particles were manipulated by changing the polarization of the beam \cite{shvedov2014long}. For asymmetric airborne particles, it was shown experimentally that by changing the intensity of a single Gaussian beam, the axial location of a trapped particle can be controlled \cite{lin2014optical}. 
Many of the beams used in optical trapping use spatial modes having Orbital Angular Momentum (OAM) \cite{allen1992orbital,padgett2000light,padgett2004light,yao2011orbital}. These modes enable both the generation of hollowed beams useful in trapping absorbing airborne particles \cite{shvedov2010optical,shvedov2009optical,shvedov2014long,shvedov2010giant} and the transportation of particles along helical trajectories \cite{lee2010optical,zhao2015curved}.

Here we present, experimentally, the use of a helical beam, made of the superposition of modes with different OAM and different axial wave vector components, to trap and transport absorbing airborne particles up- and downstream by rotating the beam one way or the other. This technique is an optical realization of the famous Archimedes' screw, used to bring water from underground wells. Not too close to the focus of the beam, the particle motion is matched to the movement of the optical screw, enabling robust and scalable two-way conveying.  We would like to note that in recent years an optical Archimedes screw was suggested for manipulating the movement of trapped cold atoms \cite{okulov2012cold,al2016rotating}.   

\section*{Results}
\subsection*{Experimental Setup}
\begin{figure}
	\centering
	\includegraphics[width=0.9\textwidth]{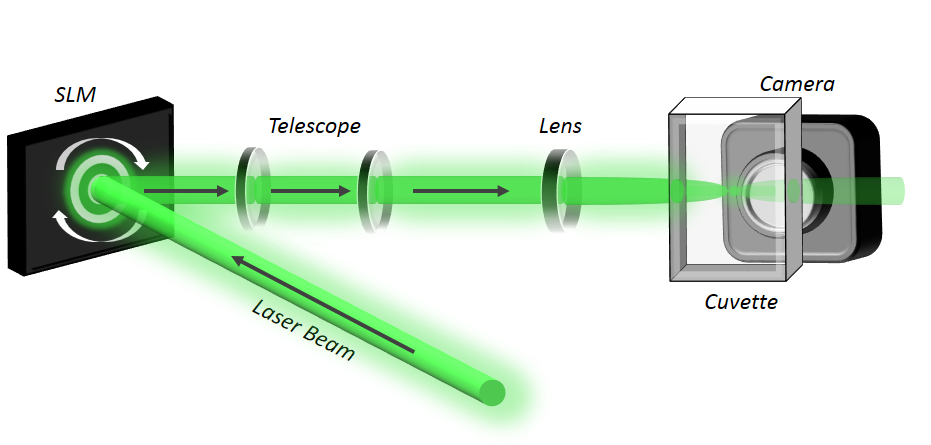}
	\caption{\textbf{Experimental Setup.} A collimated laser light reflects off of a Spatial-Light-Modulator (SLM), scaled down using a telescope, and finally focused into a cuvette filled with airborne absorbing particles. A camera records the motion of trapped particles.}
	\label{fig:setup}
\end{figure}

The experimental setup (See Fig. \ref{fig:setup}) is driven with a 532 nm CW laser (Laser Quantum Ventus 532 Solo) and uses a reflective phase only Spatial Light Modulator (Holoeye Pluto SLM). The laser beam is expanded and collimated, reflected of the SLM, magnified using a collimating telescope, and finally focused using a $5$ cm focal lens into a cuvette containing aerosol carbon nanoparticles (ALDRICH Carbon nanopowder $<$100nm particles). Observing the particles under a microscope we observe an average diameter of $~1.4$ $\mu$m, suggesting that we are actually trapping aggregates. The power of the laser after the focusing lens that was used is $53$ mW (stable trapping was observed for powers down to $\sim$30mW).  The trapped particles inside the cuvette were imaged using an LG G3 smartphone camera (13 MP, f/2.4, 29mm, phase detection/laser autofocus) and also using a  scientific CMOS camera (Ophir Spiricon SP620U Beam Profiling Camera) on a translation stage aligned parallel to optical axis of the beam. This same camera was also used together with a magnifying imaging lens (focal length 5cm), on a translation stage, in front of the beam with various filters to capture the beam profile together with a trapped particle while a 633nm laser (with intensity on the trapped particle much smaller than that of the trapping laser) illuminates the trapped particle from the side. Different phase patterns with relative rotation of $2\pi/M$ (with $M$=50), were used to rotate the beam. The phase patterns were projected sequentially by the SLM, at a controlled rate, setting the beam's rotation velocity. Particles motion was extracted from raw film footage by first increasing the image contrast and then using the Hough transform \cite{illingworth1988survey} for feature extraction.

\subsection*{Construction of the optical screw}

The optical Archimedes' screw is a standing-wave with a helical intensity profile. There are several works showing that a helical intensity profile can be generated by superposing modes with different OAM and different linear momenta \cite{schulze2015accelerated,vetter2014generalized,kotlyar2007rotation,rop2012measuring,schechner1996wave,tao2006residue,vasilyeu2009generating,zheng2015rotating,dudley2012stationary}. Similarly, in our case, the optical screw is constructed from a superposition of two beams carrying different linear and angular momentum. Basically, a superposition of modes characterized with different momenta results in a standing-wave in the conjugate coordinate. This is true whether the conjugate pair is linear momentum and linear coordinate or angular momentum and azimuthal coordinate. Thus, the superposition of two spatial modes which differ in both their linear and angular momenta would result in a standing-wave pattern in both the longitudinal and azimuthal directions with the overall effect of realizing a helical intensity profile. The modes we use are based on ideal Bessel beams described with \cite{durnin1987exact,mcgloin2005bessel}
$E(r,\phi,z)=A_0 e^{ik_z z} J_n (k_r r) e^{\pm in\phi}$,
%
%
%
%
%
where $A_0$ is an arbitrary field amplitude, $ k_z $ and $ k_r $ are the wave number components in the longitudinal and transverse directions, respectively, $ J_n(x) $ is the Bessel function of the $n$-th order, and $ n $ is a dimensionless integer (known as the topological charge) representing the orbital angular momentum  quantized in units of $\hbar$ per photon \cite{padgett2000light,padgett2004light,yao2011orbital}. When $n>0$ such a mode is known as a vortex Bessel beam.

\begin{figure*}
	\centering
	\includegraphics[clip, trim=0.4cm 6.5cm 2cm 5.5cm,width=1.0\textwidth]{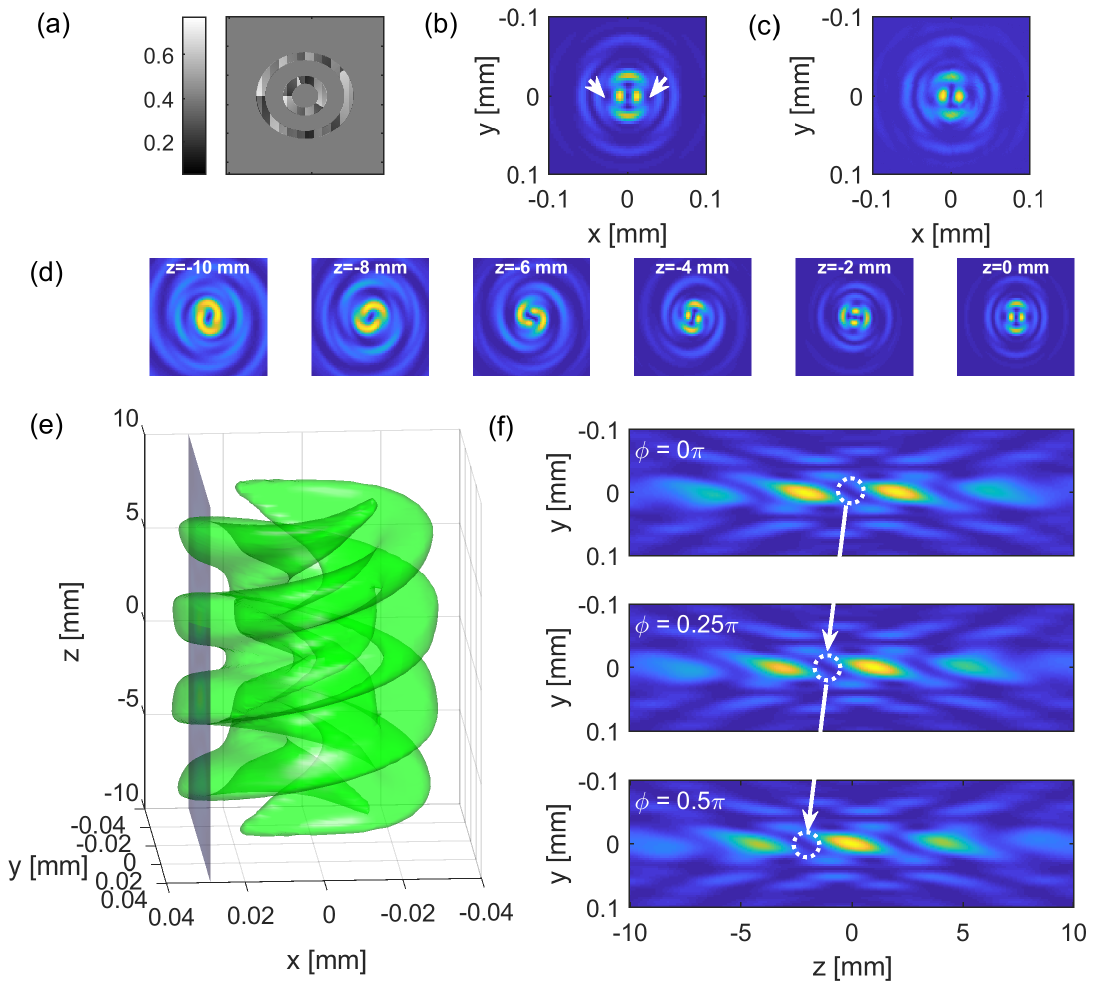}
	\caption{\textbf{The Optical Archimedes' screw.} (a) The Fourier domain mask that was applied to the phase-only SLM. (b) Calculation of the beam profile in the focal plane at the middle of the cuvette. The arrows indicate dark volumes in which particles were observed to be trapped. (c) The beam profile captured on a camera, using a longer focal lens than used for trapping, and scale adjusted for the lens used for trapping. (d) Calculation of the beam profile in several locations along the propagation length. (e) Simulation of several iso-intensity surfaces of the beam around the focal plane. (f) Simulations, for several rotation angles $\phi$ of the optical screw, of a plane cut (as shown in (e)) of the beam parallel to the optical axis and displaced transversely to the point indicated by one of the arrows on panel (b) above. The dashed white circles denote a possible trap location for the particle. The dashed arrows are a guide to the eye.}
	\label{fig:beamprofile}
\end{figure*}

The spectral representation of such an ideal beam is an infinitely thin ring of uniform intensity and an azimuthally linear phase profile with an $n$-th order rotational symmetry \cite{durnin1987diffraction}. The radius of the ring sets the value of the longitudinal linear momentum. In our case, we create a superposition of two non-ideal vortex Bessel beams in Fourier space by using two finite-thickness rings, with different radii and opposing phase profiles, characterized with $n=\pm 1$. We then multiply the phase pattern with a blazed phase grating so that only light which is reflected off the rings area is directed into the 1st diffraction order of the grating. To realize this pattern physically, we use a phase-only SLM in a 2f configuration.  The phase pattern realized on the SLM is shown in Fig.~\ref{fig:beamprofile}(a). The measured beam cross section after magnification and 2f Fourier transform at the focal plane of the lens agrees well with its expectation (compare Fig.~\ref{fig:beamprofile}(b) and (c)). Its evolution along the axial direction is shown in Fig.~\ref{fig:beamprofile}(d). We note, that the experimentally measured beam cross section was taken using a longer focal lens to improve resolution.  A calculation, using Fresnel field propagation, shows iso-intensity surfaces of the beam in a volume around the focal plane (see Fig.~\ref{fig:beamprofile}(e)). Here, the pitch of the screw is half its axial period, while generally, for a similar beam constructed from two modes with charges of $\pm n$, the ratio between the period and pitch is $2n:1$. The condition with which the pitch $\Lambda$ can be calculated is $I(r,\phi,z+\Lambda)=I(r,\phi,z)$ where $I$ is the intensity of the beam. The condition for calculating the period $\Delta z$ is $I(r,\phi,z+\Delta z)=I(r,\phi+2\pi,z)$.  The pitch is fixed by the radii of the rings on the SLM and is given by:
$\Lambda=\lambda \cdot [cos(\theta_1)-cos(\theta_2)]^{-1}$,
%
%
where $\theta_i=tan^{-1}(\frac{R_i}{f})$, $R_i$ is the radius of each ring on the mask and $f$ is the focal length of the lens. Switching the sign of the topological charge associated with each ring would switch the helicity of the whole beam.

Each of the vortex Bessel beams is approximately non-diffracting along a length of $L_{ND}={2\lambda f^2}/(WR)$ \cite{lobachinsky2017fly},
%
%
%
where $W$ is the width of the ring and $R$ is its radius. Along this length, the optical screw keeps its form, while further up or downstream it starts to broaden due to diffraction. The central part of the beam is composed of $2n$ intertwining strands. A rotation of the optical screw around the propagation axis is easily accomplished by rotating either one or both of the rings on the SLM. In the first case, the interference pattern is rotating, in the second, the whole beam. The overall effect is similar, with the difference that a $\Delta \theta$ rotation of one of the rings is the same as rotating the whole beam by $\Delta \theta /2$.  When a particle is trapped at one of the off-axis dark volumes in the beam, it is carried upstream or downstream, depending on the direction of the beam rotation and on the beam helicity (which is constant in our experiment). 


\subsection*{Screw-velocity-matched particle motion}
A particle trapped in the optical screw maintains its position as long as the screw is stationary. The trapping stiffness is estimated to be around $50$ pN/m (see the Supplemental section below). When the screw is rotating, the trapped particle is translated in a controlled manner, as long as it remains far enough from the focal plane of the beam ($1$ mm away in our case), and is driven at velocities lower than $\sim 0.3$mm/sec. The axial movement of the particle in these cases is close to the axial phase velocity of the optical screw, defined as the screw period divided by the screw's rotation time period. The particle is repelled from the high-intensity regions of the rotating  beam, very similarly to water or air molecules being repelled from the blades of a rotating fan. 
The small diffraction of the screw is responsible to the difference between the actual particle axial velocity to the calculated phase velocity (which is exact only for an infinitely non diffracting beam). Still we describe this type of movement as screw-velocity-matched.

\begin{figure}
		\centering
	\includegraphics [clip, trim=1.3cm 0cm 2.2cm 0cm,width=0.8\textwidth]{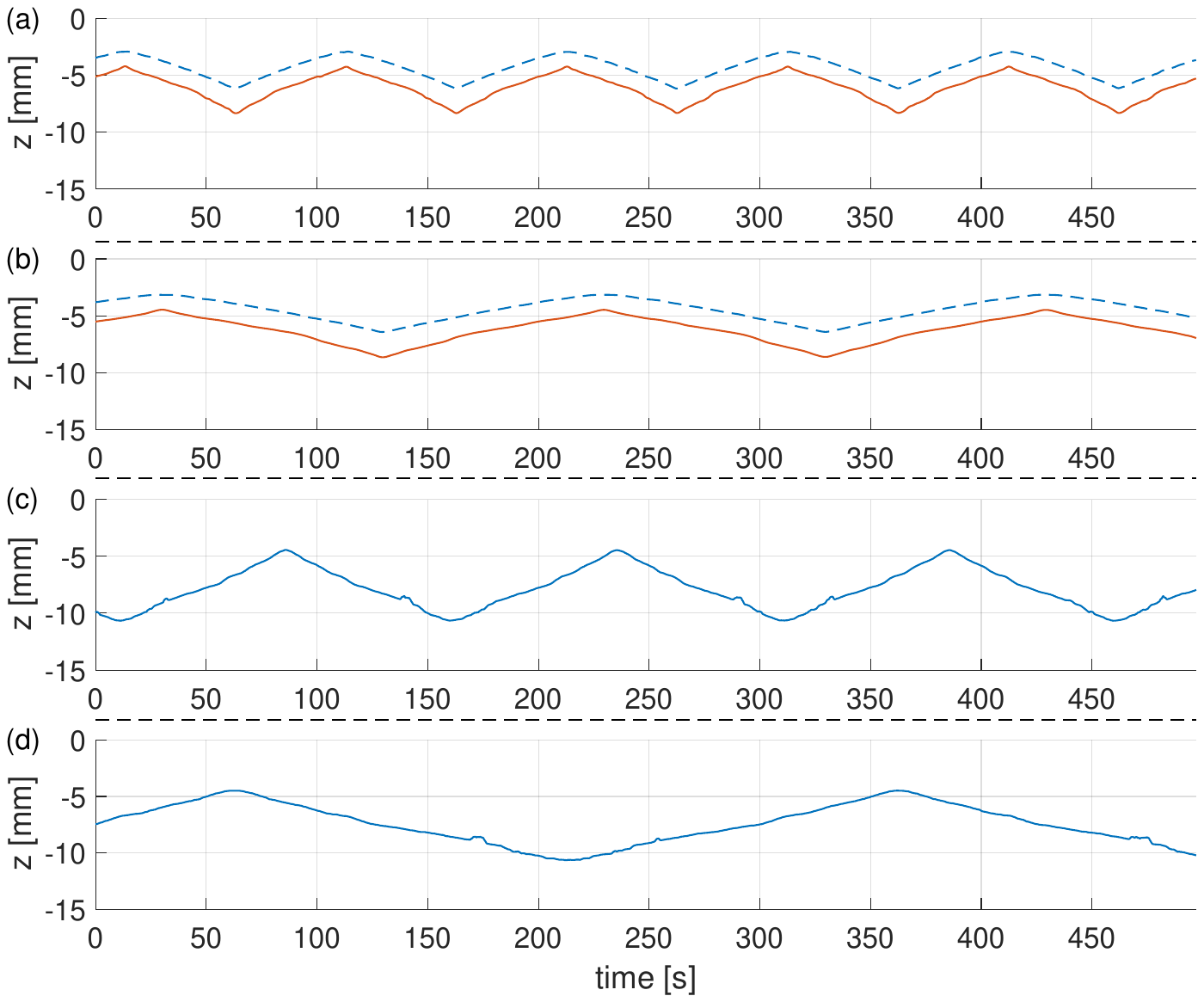}
	\caption{\textbf{Screw-velocity-matched particle axial motion.} In all cases the optical screw is periodically switching its rotation direction after $N$ rotations and its axial phase velocity is $v_p$. (a)  $N=0.5$, $v_p= 0.08$ mm/sec. (b)  $N=0.5$, $v_p= 0.04$ mm/sec. (c)  $N=0.75$, $v_p= 0.08$ mm/sec. (d)  $N=0.75$, $v_p= 0.04$ mm/sec. Two particles were trapped together in cases (a) and (b).}
	\label{fig:periodicalmotion}
\end{figure}

As a demonstration we accomplish periodic oscillatory motion of trapped particles, by repeatedly rotating the screw one way, and then the other. This oscillatory motion is repeated with two different rotation speeds and with different number of rotations of the screw to each side. The axial ($z$ direction)  movement of the trapped particles for these cases are given in Fig. \ref{fig:periodicalmotion}.
The two phase velocities we used are $0.08$ mm/sec (Fig.\ref{fig:periodicalmotion}(a),(c)) and $0.04$ mm/sec (Fig. \ref{fig:periodicalmotion}(b),(d)). The number of screw rotations we used to each side was half a turn ($N=0.5$, Fig. \ref{fig:periodicalmotion}(a),(b)) and three quarters of a turn ($N=0.75$, Fig. \ref{fig:periodicalmotion}(c),(d)). In total, we have then four different movements of the screw and the trapped particle is locked and velocity matched to this movement. 
The pitch of the screw in all cases was $4.1$ mm (The optical screw parameters were chosen to create about 2 full turns of the screw within the confines of the cuvette whose length along the optical axis is 2 cm).  Incidentally, for the cases shown in Fig. \ref{fig:periodicalmotion}(a),(b), two particles were trapped at the same time at two different axial locations of the screw (separated by $2$ mm). It can be seen that they move in tandem performing similar motion. Using a magnifying imaging system, it was verified (with an error of $\pm 0.44$$\mu m$) that the particles do not move along the transverse directions, while their location within the beam profile is set off-axis as indicated using an arrow in Fig. \ref{fig:beamprofile}(b).  As the particles are transversely displaced off-axis they experience an intensity landscape that moves axially as the screw is being rotated. This can be appreciated by examining for different rotations of the optical screw,  a plane cut of the beam whose location is parallel to the optical axis and displaced off-axis to the position of the trapped particles (see Fig. \ref{fig:beamprofile}(f)).   

\begin{figure}
	\centering
	\includegraphics [clip, trim=3.5cm 3cm 2.6cm 1cm,width=0.8\textwidth]{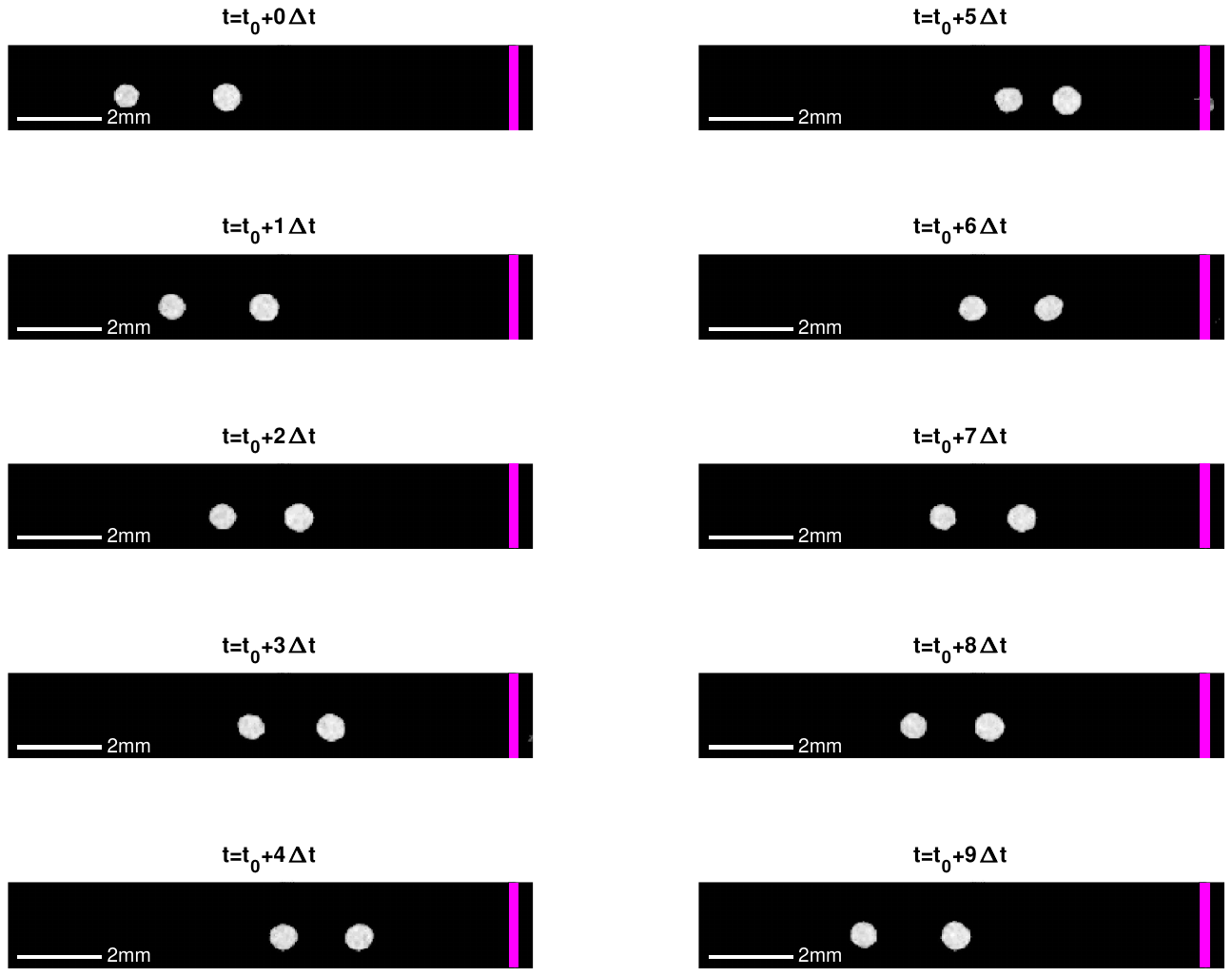}

	\caption{\textbf{Camera snapshots during the oscillatory movement of two particles trapped together in a rotating optical screw.} During the interval the snapshots were taken, the optical screw is first rotating to move the particles upstream (left) and then downstream (right). The optical screw phase velocity is $v_p= 0.08$ mm/sec. The two particles are trapped at an average distance of $2$ mm from each other. The focal plane is marked with the vertical (purple) line on the right hand side of each image. $t_0$ was set to a moment just after a change of direction of the particles movement. $\Delta t=10$ s. } 
	\label{fig:framepropogation}
\end{figure}

We would like to note that as the beam retains some slow divergence (it is not perfectly non-diffracting), then the axial velocity of conveyed particles is slightly dependent on the axial position of the particles. This can be seen for the two particles moving in tandem in Fig.~\ref{fig:periodicalmotion}(a),(b), where due to their different axial position, their axial velocities are also slightly  different and so the distance between them changes along their trajectories. This can be seen more clearly in Fig.~\ref{fig:framepropogation} showing camera snapshots for the pair of particles during an oscillation of their movement.   


\subsection*{Periodic particle motion with non-matched velocities}

\begin{figure}
	\centering
	\includegraphics [clip, trim=1.3cm 0cm 2.2cm 0cm,width=0.8\textwidth]{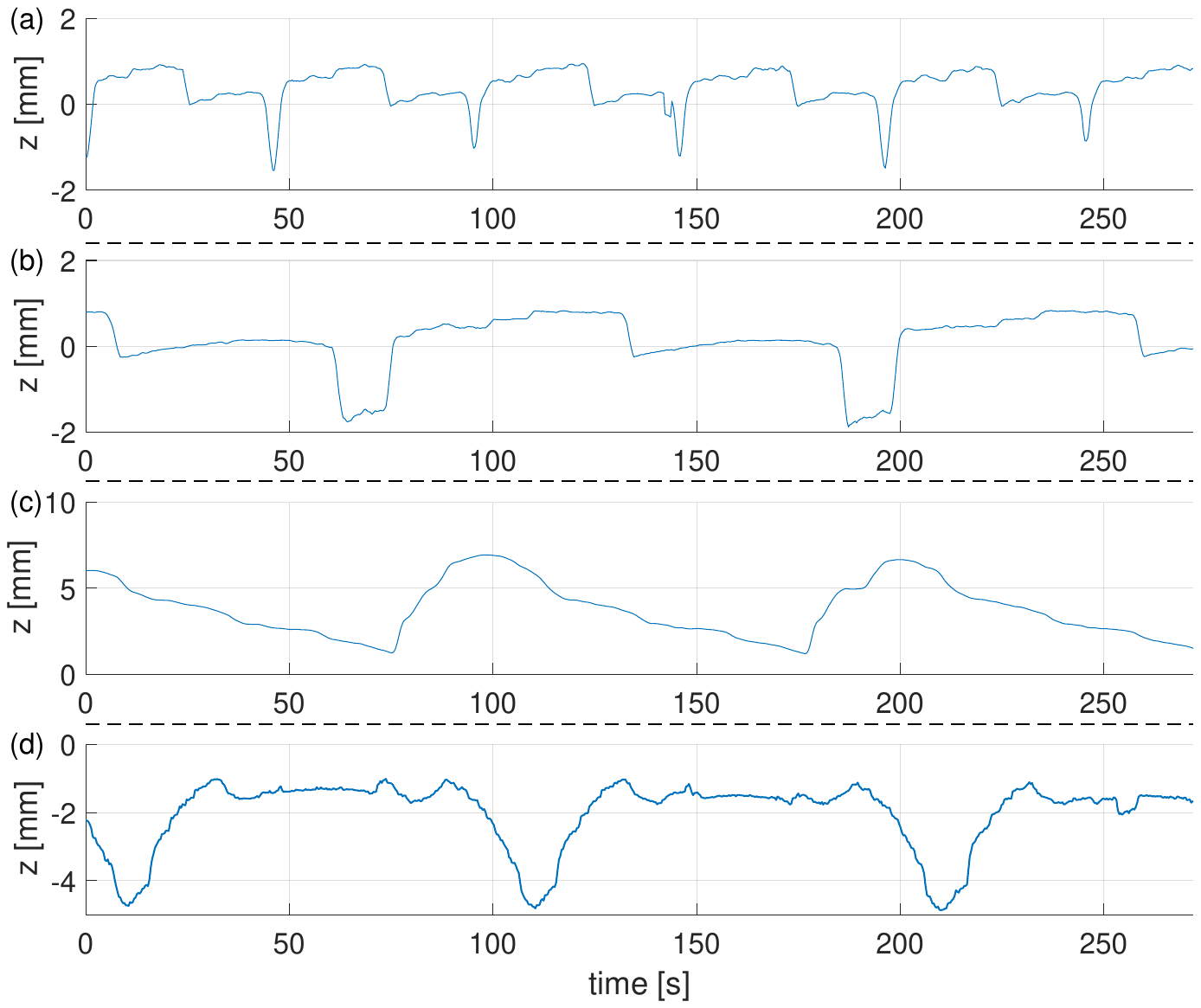}
	\caption{\textbf{Non velocity-matched periodic particle axial motion.} (a) The optical screw is constantly rotating to deliver particles downstream at phase velocity of $v_p=0.08$ mm/sec and pitch of $4.1$ mm. (b) downstream, $v_p=0.03$ mm/sec, pitch is $4.1$ mm (c) upstream, $v_p=0.08$ mm/sec, pitch $4.1$ mm (d) The optical screw rotation direction is switched every half a turn (every pitch cycle), $v_p=0.23 $ mm/sec, pitch of $11.3$ mm. } 
	\label{fig:nearfocusmotion}
\end{figure}

When particles are brought by the rotation of the optical screw close to the focal plane of the beam (in our case, about $1$ mm from the focus) their velocities do not continuously match the phase velocity of the optical screw, and their motions exhibit complex periodic behavior.
In such cases, the nature of the motion is sensitive both to the rotation dynamics of the optical screw, and to the location of the particle with respect to the focal plane. However, all motions we have tracked exhibit periodic oscillatory behavior, regardless if the screw rotation is constant or switching directions.  In Fig.~\ref{fig:nearfocusmotion}(a) and (b), the screw was rotating continuously with the beam direction (striving to deliver particles downstream) at velocities of $0.08$ mm/sec and $0.03$ mm/sec, respectively, with a pitch of $4.1$ mm. In both cases, the particle crossed the focal plane of the beam, lingering for a relatively large portion of its period very close to the focus. Eventually it was pushed back and then carried again towards the focus, where it lingered again and so on. In Fig.~\ref{fig:nearfocusmotion}(c) the screw was rotating continuously against the beam direction with a velocity of $0.08$ mm/sec and a pitch of $4.1$ mm, aiming to move the particle upstream. In this case, the particle was unable to cross the focal plane. While it was being carried towards the focus it moved together with the screw phase velocity, but once it arrived close to the focal plane it was pushed, relatively rapidly, downstream, up to a point where the optical field of the screw was able to trap and carry it again upstream. 

Finally, in Fig.~\ref{fig:nearfocusmotion}(d) the optical screw motion is oscillatory - rotating half a turn to each  side at a phase velocity of $0.23$ mm/sec with a pitch of $11.3$mm. In this case, the particle is carried towards or away from the focal plane at a velocity of $0.22$ mm/sec locked to the phase velocity of the optical screw. However, when it comes close to the focal plane it lingers a long time at a specific axial position. The particle is able to escape only when the screw force is directed to pull it out, upstream.  In all cases described here the motion of the trapped particle is no longer velocity matched at all times to the motion of the optical screw. However, the  motion is still locked to the rotation of the optical screw in the sense that the periodicities of all the various motions described here are equal to that of the optical screw. Namely, the period of the motion of the particles are equal to the rotation period of the optical screw in the cases where the screw is rotating continuously to the same direction, and to twice this value when the screw is changing its rotation direction after each full revolution. 

We note, that the periodic motions observed here are reminiscent of a somewhat similar phenomenon observed lately due to the interplay of photophoretic and direct optical forces \cite{lu2017light}. However, here the motion is on a much longer scale, and is driven by changing the intensity profile (by rotation) of the light beam.

\section*{Conclusions and Discussion}

To conclude, we have experimentally demonstrated the transfer of rotational movement of a helical beam to the axial movement of a trapped particle within the beam, by realizing an optical analogue to the famous Archimedes' screw. We observe two regimes of motion depending on the proximity of the trapped particle to the focal plane of the beam. 
If the trapped particle is located far enough from the focal plane, the motion of the particle is velocity matched to the movement of the optical screw, whether the axial velocity of the screw is directed up or downstream. If the particle is close enough to the focal plane it exhibits periodic motion locked to the period of the screw rotation, however its motion is no longer velocity match at all times to the motion of the optical screw. The axial amplitude of the motion demonstrated in our setup was about $0.5$ cm. However, the simplicity of the method and its basic operation principle suggest that it could be extended either to longer scales, or to smaller dimensions with better resolution. Furthermore, in contrast to other methods of conveying airborne particles \cite{shvedov2010giant,shvedov2014long}, where the velocity of the particles is dictated by the light intensity and the dimensions of the particles, here the velocity is dependent only on the rotation speed of the optical screw. It is an interesting question whether this method could be applied to weakly absorbing particles in suspension. In cases where the particles are attracted to light, rather than repelled from it, we expect that the particles would stick to the helical beam in a given transverse plane, and so if possible at all, a more elaborate scheme would probably be needed to transfer the particles axially while the helical beam is rotated. Finally, we expect this method to be useful for applications with multiple particles, for example, for the construction of an optical turbine.

\section*{Funding Information}
H.N, T.A and Y.R wish to acknowledge support from the ISF, grant No. 899/17.

\section*{Supplementary section- Estimating the stiffness of the optical trap}
While being conveyed by the rotating screw, the particles do not move on average along the transverse direction. However, they do undergo random, noisy, fluctuations in their position along this direction which is characterized by an average distance of $8$ $\mu$m  from the mean location with variance of about $114$ $ \mu m ^2$ and maximum amplitude of $25$ $\mu$m . Using a standard calculation which fits the transverse noise statistics to a thermal Boltzmann distribution \cite{neuman2004optical} the stiffness of the optical trap in the transverse direction can be calculated. The results set the average stiffness observed, for laser trapping power of $53$ mW, at room temperature, to be  around $50$ pN/m. Using the average distance of the particles from the equilibrium point ($8$ $\mu$m)  we calculate an average radial force exerted by the trap of an $F_r=0.4$ fN. We note that the expected radiation force acting on the particles: $F_{rp}=6.6 \cdot 10^{-4}$ fN (calculated using the model appearing in Ref. \cite{desyatnikov2009photophoretic}) is much smaller, while gravity acts with: $F_g=0.14$ fN, both calculated for an aggregate particle size of $1.4$ $\mu$m (as observed under a microscope outside the trapping setup). 

We can compare these numbers to an analytical model, specifically using Eq. 2 in Ref. \cite{shvedov2010optical}, where we replace the distance of the trapped particle from the center of a vortex beam (as was used in that work) with the average distance of the particles movement within the dark volume at which they are trapped in our case ($8$ $\mu$m), and we replace the inner vortex radius with an estimated radius averaged over a few dark volumes in our beam ($29$ $\mu$m, which is similar to the maximum observed noise amplitude of the particles' transverse motion of $25$ $\mu$m). Particles density and thermal conductivity were taken to be the same as in Ref. \cite{shvedov2010optical}. With these, the model agrees with our stiffness measurement when the  particle (aggregate) diameter used is $200$ nm. When an aggregate average particle diameter of $1.4$ $\mu$m   is used in this model the stiffness parameter is much higher: $30$ nN/m, leading to stronger radial force exerted by the trap: $F_r=0.24$ pN. This model is highly sensitive to the particle size and to the size of the dark volume at which the particle is confined.



%
%
%

\bibliography{bibfile}



\end{document}